\documentstyle[aps,epsf,rotate,multicol]{revtex}
\title{Excitation spectrum of Andreev billiards with a mixed phase space}
\author{H. Schomerus and C. W. J. Beenakker}
\address{Instituut-Lorentz, Leiden University, P.\,O.~Box 9506, 2300 RA Leiden,
  The Netherlands}

\newcommand{\bcols}{\ifpreprintsty\else\begin{multicols}{2}\fi}
\newcommand{\ecols}{\ifpreprintsty\else\end{multicols}\fi}
\begin{document}
\draft
\date{October 1998}
\maketitle

\begin{abstract}
We present a semiclassical theory for the
excitation spectrum of a ballistic quantum dot weakly
coupled to a superconductor, for the
generic situation that
the classical motion gives rise to a
phase space containing islands of regularity in a chaotic sea.
The density of low-energy excitations  is
determined by
quantum energy scales that are related in a simple way
to the morphology of the mixed phase space. 
An exact quantum mechanical computation
for the annular billiard shows good agreement with the
semiclassical predictions, in particular for the reduction of the
excitation gap when the coupling to the regular regions is maximal.
\end{abstract}
\bcols

The spectral statistics of quantum systems is intimately related to
the nature of the corresponding classical dynamics
\cite{Gutzwiller:Haake}.
Two celebrated examples are that 
chaoticity of the classical dynamics is reflected in the quantum realm
by level repulsion while 
integrability causes level clustering
\cite{Berry:Tabor}.
Recently, confined two-dimensional electron gases
(quantum dots) coupled to a superconductor via a ballistic point
contact
have become a new arena
for the study of 
quantum-classical correspondences \cite{Melsen,Altland,Lodder,Altland:Simons}. 
Such systems are commonly called Andreev billiards \cite{Kostzin},
because of the alternation of ballistic motion (as in a conventional billiard)
with Andreev reflection \cite{Andreev} at the interface with the
superconductor.
The proximity of the superconductor causes a depletion of excited
states at low energies (proximity effect).
It was found \cite{Melsen} that a chaotic Andreev billiard has an
excitation gap of the order of the Thouless energy,
while an integrable Andreev billiard has no true gap but an
approximately linearly
vanishing density of states.
(The Thouless energy $E_T=g\delta/4\pi$ is the product of the point contact
conductance $g$, in units of $2e^2/h$, and the mean level spacing
$\delta$ of the isolated billiard.)

Both chaotic and integrable dynamics are atypical. 
The generic situation is a mixed phase space,
with ``islands'' of regularity separated from 
chaotic ``seas'' by
impenetrable dynamical barriers.
A generally applicable theory for the proximity effect
in ballistic systems should address the case of a mixed phase
space. In this paper we present such a theory.

\begin{figure}
\epsfxsize7cm
\hspace*{.4cm}\epsffile{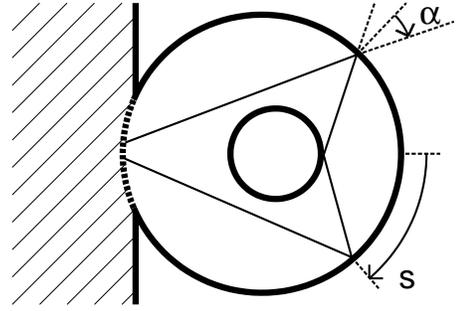}
\caption{Andreev billiard consisting of a confined normal
\narrowtext
conducting region interfacing with a superconductor (shaded)
over a distance $W$. The normal region is shaped like an annular
billiard, bounded by two excentric circles (outer radius $R$,
inner radius $r$, distance of origins $\rho$).  This figure represents
the case $R=1$, $r=0.35$, $\rho=0.1$, $W=0.8$.  A periodic
trajectory is indicated, involving two Andreev reflections at the
interface.  For the Poincar{\'e} map one monitors the collisions
with the outer boundary (angle of incidence $\alpha$ and coordinate
$s$ along the boundary, with $\alpha=0$ denoting normal incidence
and $s=0$ denoting the point closest to the inner circle).
}
\label{fig1}
\end{figure}

In a semiclassical approach we link
the excitation spectrum quantitatively and qualitatively to the morphology of
non-communicating regions in phase space. Different regions exhibit
greatly varying length scales, which also depend sensitively on the
position of the point contact. Still, we find 
that a general
relation exists (in terms of effective Thouless energies) between
these classical length scales and
the corresponding quantum energy scales.
The results for integrable and fully chaotic motion are recovered as
special cases.
For the mixed phase space our main finding is a reduction
of the excitation gap below the value $E_T$ of fully chaotic
systems.
The reduction can be an order of magnitude, as we illustrate
by a numerical calculation in the annular billiard
\cite{Bohigas:Boose:1993} shown in Fig.\ \ref{fig1}.

We consider a two-dimensional ballistic region (a ``billiard'')
of area
$A$ (mean level spacing $\delta=2\pi \hbar^2/mA$)
that is connected to a superconductor by an opening of width $W$
(corresponding to a dimensionless conductance $g=2W/\lambda_F$,
where $\lambda_F$ is the Fermi wavelength). Classical trajectories
consist of straight lines inside the billiard, with specular
reflections at the boundaries and retro-reflections ($=$ Andreev
reflections) at the interface with the superconductor. 
We assume that $\delta \ll E_T \ll \Delta$, where $\Delta$ is the
excitation gap in the bulk superconductor. The first condition,
$\delta \ll E_T$ or $W\gg \lambda_F$, is required for a
semiclassical treatment. The second condition, $E_T \ll \Delta$,
ensures that the excitation spectrum becomes independent of the
properties of the superconductor.

The Andreev billiard has a discrete spectrum for $\varepsilon <
\Delta$.
(The excitation energy $\varepsilon >0$ is measured with respect to the
Fermi energy. We count each spin-degenerate level once.)
For $\varepsilon \ll \Delta$ the semiclassical expression
for the density of states $\rho(\varepsilon)$ reads \cite{Melsen}
\begin{equation}
\label{eq:scquant}
\rho(\varepsilon)= \frac2\delta\int_0^\infty \!\!\!\!{\rm d}L\,P(L)
\sum_{n=0}^\infty\delta\left(\frac{\varepsilon}{2 \pi E_T}
-\left(n+\case 12 \right)\frac {L_T}{L}
\right)\;,
\end{equation}
with $P(L)$ the distribution of path lengths between subsequent Andreev
reflections. The distribution $P(L)$ is normalised to unity and based on
a measure ${\rm d}s\,{\rm d}\sin \alpha$ for the initial conditions at
the interface with the superconductor (see Fig.\ \ref{fig1}).
The length scale $L_T\equiv \hbar v_F/2 E_T$  
(with $v_F$ the Fermi velocity) 
is determined by the geometry of the billiard
by $L_T=\pi A/W$ and is therefore purely classical.
Eq.\ (\ref{eq:scquant}) follows directly from the Bohr-Sommerfeld
quantisation rule \cite{Melsen}. The derivation of Lodder and Nazarov
\cite{Lodder} starts from the Eilenberger equation \cite{Eilenberger}
for the quasiclassical
Green function and arrives at an expression for $\rho(\varepsilon)$ 
that is mathematically equivalent to Eq.\ (\ref{eq:scquant}).

Return probabilities like $P(L)$ and the related decay of classical correlations
have been addressed in many studies \cite{Bauer}.
In a chaotic billiard, $L_T$ is the mean path length and $P(L)
\propto
\exp (-L/L_T)$ is an exponential distribution.
Eq.\ (\ref{eq:scquant}) then
gives the density of states
\begin{equation}
\label{eq:chaoticdos}
\rho(\varepsilon)=\frac {2 x^2} \delta \frac{\cosh x}{\sinh^2 x}
\;,\quad x=\frac{\pi E_T}{\varepsilon}\;,
\end{equation}
which drops from $2/\delta$ (the factor of two arises because
both electron and hole excitations contribute at positive $\varepsilon$)
to exponentially small values as $\varepsilon$
drops below $\approx 0.5\, E_T$.
Eq.\ (\ref{eq:chaoticdos}) is close to the exact
quantum mechanical result \cite{Melsen}, which has $\rho\equiv 0$ for
$\varepsilon\le 0.6\, E_T$. 
For integrable motion $P(L)$ decays algebraically $\propto L^{-p}$
with $p$ close to 3. Eq.\ (\ref{eq:scquant}) then gives $\rho(\varepsilon)
\propto \varepsilon^{p-2}$, hence an approximately
linearly vanishing density of states.
Numerical studies on the circular and
rectangular billiard confirm the validity of the semiclassical approach
\cite{Melsen,numerics}.

The semiclassical expression (\ref{eq:scquant}) holds also for mixed dynamics.
It allows
to regard each non-communicating region in phase space as a
distinct system, to be labelled by an index $i$.
It is helpful to rewrite Eq.\
(\ref{eq:scquant}) in terms of an effective level spacing $\delta_i$
and Thouless energy $E_{T,i}\equiv \hbar v_F/2 L_{T,i}$
for each of these regions.
(This approach
extends the Berry-Robnik conjecture \cite{Berry:Robnik} to open systems.)
We decompose 
$\rho=\sum_i \rho_i$ into partial densities of states $\rho_i$, defined by
\begin{equation}
\label{eq:scquantmixed}
\rho_i(\varepsilon)= \frac 2{\delta_i}\int_0^\infty \!\!\!{\rm d}L\, P_i(L)
\sum_{n=0}^\infty\delta\left(\frac{\varepsilon}{2 \pi E_{T,i}}
-\left(n+\case 12 \right)\frac {L_{T,i}}{L}
\right)
\;.
\end{equation}
The distribution $P_i(L)$ (still normalised to unity) now pertains to
initial conditions (still with measure ${\rm d} s\,{\rm d}\sin\alpha$)
on the interface with the superconductor that evolve into the $i$-th
region of phase space.
On the scale $L_{T,i}$, the distribution
$P_i(L)$ decays exponentially for
chaotic parts of phase space while algebraic decay is found for regular
regions \cite{tails}. 
In each case the partial density of states $\rho_i$ rises to a value
$2/\delta_i$ on an energy scale $E_{T,i}$,
but while $\rho_i$ has an
excitation gap for the chaotic regions it rises linearly for the regular
regions.
Eq.\ (\ref{eq:scquantmixed})
applies to those regions
that are accessible for a given location of the interface.
We call these ``connected'' regions.
The other ``disconnected'' regions (usually some of the regular islands)
do not feel the proximity of the superconductor
and give a constant background contribution 
$\rho_i(\varepsilon)= 2/\delta_i$ in the semiclassical
approximation.

\begin{figure}
\epsfxsize8cm
\epsffile{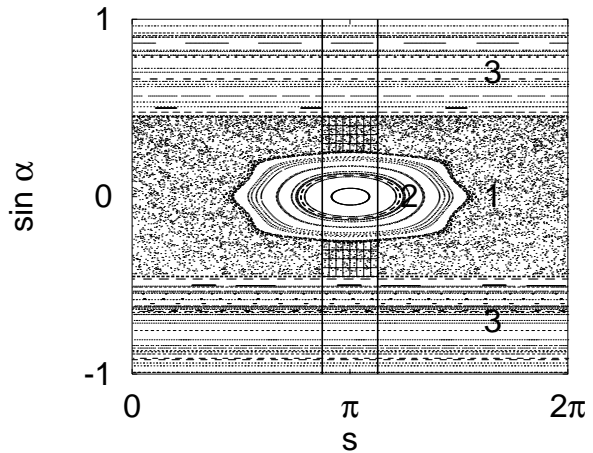}
\caption{
\narrowtext
Poincar{\'e} map
of the annular billiard of Fig.\ \protect\ref{fig1}.  
Dynamical barriers separating regions in phase space are shown
as dashed lines.  Chaotic trajectories are found in region 1.
Region 2 is an island of regular motion around a short stable
periodic orbit.  Region 3 is integrable and consists of skipping
orbits that never hit the inner circle. Initial conditions at the
interface with the superconductor are uniformly distributed
within the strip around $s=\pi$, of area $2W$. The hatched area
is the overlap $O_1$ of this strip with region 1.
}
\label{fig2}
\end{figure}

Two phase space measures $O_i$ and $V_i$
determine the mean length $L_{T,i}=V_i/O_i$ between Andreev
reflections,
the effective level spacing $\delta_i=(2\pi\hbar)^2/mV_i$,
and the effective Thouless energy
$E_{T,i}=\hbar v_F/2 L_{T,i}=g_i\delta_i/4\pi$, where
$g_i=O_i/\lambda_F$ is the effective dimensionless conductance.
The first is the area $O_i$ that the region overlaps with the
superconducting interface
on the Poincar{\'e} map
(see Fig.\ \ref{fig2}). It is a measure of the coupling strength of a region
to the superconductor. 
The second is the volume $V_i$
that the region fills out in the full phase space $({\bf r}, \phi)$,
where the coordinate $\bf r$ extends over the area of the billiard and
$\phi\in [0,2\pi)$ is the direction of momentum.

The phase space that is explored from the point contact can also be
parameterised by the variables
$s$, $\sin\alpha$ on the interface and a coordinate $l$ along  the
trajectory. The identification of
$L_{T,i}=V_i/O_i$ as the mean path length in region $i$ is
a consequence of
${\rm d} s\,{\rm d}\sin\alpha \,{\rm d}l={\rm d}{\bf r}\,{\rm d}\phi$.
The mean length 
of all trajectories
$\langle L \rangle\equiv\int {\rm d}L\,L P(L)=
{\sum_i}^\prime O_i L_{T,i}/2W = {\sum_i}^\prime V_i/2W$
can be used
to determine the total phase space volume
$V_{\rm con}\equiv
{\sum_i}^\prime V_i=2 W\langle L\rangle$
that is connected to the interface with the superconductor. Here the
prime denotes restriction of the
sum to connected regions, and we used the sum rule ${\sum_i}^\prime O_i=2W$.
The volume $V_{\rm dis}$
of the disconnected regions (which determines the background
contribution to $\rho$) follows from the sum rule $\sum_i V_i=V_{\rm
con}+V_{\rm dis}=2\pi A$.

Since typically
the smallest $E_{T,i}\ll E_T$,
the total density of states $\rho=\sum_i\rho_i$
has a reduced excitation gap.
This is especially the case
when one couples maximally to the regular regions.
Then their contribution
to $\rho$ at small $\varepsilon$ (long path
lengths) is minimal, and
the gap is substantially reduced 
due to long chaotic trajectories.
The constant background and the linear increase from regular regions
dominates
when the coupling is mainly to the chaotic parts of phase space.

The preceding paragraph summarises the key finding of our work.
We illustrate it now for the annular billiard of Fig.\
\ref{fig1}.
The Poincar{\'e} map in Fig.\ \ref{fig2}
shows three main regions \cite{subdivision}, one with chaotic motion (1)
and two with regular motion (2 and 3).
The regular island 2 corresponds to orbits that
bounce back and forth between the two circles
where their distance is largest. It has a short stable periodic orbit in
its centre.
Region 3 is integrable and consists of skipping orbits
(trajectories that do not hit the inner circle, so that their
angular momentum $\sin\alpha$ is conserved).

\begin{figure}
\epsfxsize8cm
\center{\epsffile{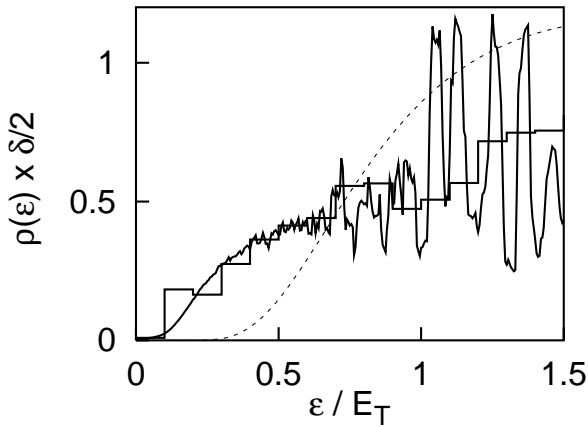}}
\caption{
\narrowtext
Density of states of the annular billiard of Fig.\ \protect\ref{fig1}.
The solid curve is the semiclassical prediction computed from
Eq.\ (\protect\ref{eq:scquant}).  The histogram is obtained by an exact quantum
mechanical computation.  The dashed curve is the semiclassical
result (\protect\ref{eq:chaoticdos}) for completely chaotic dynamics.  
}
\label{fig3}
\end{figure}

The regular regions couple maximally to
the point contact when it is located at the short stable periodic orbit,
as in Fig.\ \ref{fig1} (location $s=\pi$).
We have computed $P(L)$ by following trajectories and obtained
$\rho(\varepsilon)$ from Eq.\ (\ref{eq:scquant}). The result is shown in
Fig.\ \ref{fig3} (solid curve).
We see an excitation gap which is about a factor of 4 smaller than in
the fully chaotic case [Eq.\ (\ref{eq:chaoticdos}), dashed curve].
The reduction originates from long chaotic trajectories
with mean length $L_{T,1}\approx 4 L_T$, hence
$E_{T,1}\approx E_T/4$.
An exact quantum mechanical calculation 
\cite{numerics2} (histogram) confirms the low-$\varepsilon$
behaviour found semiclassically.
The sharp peaks at higher $\varepsilon$ in the semiclassical prediction,
which arise from families of regular trajectories of almost identical length,
are not resolved in the histogram.
This is not a surprise since
numerically accessible Fermi wavelengths
are still larger than the extension of the families.

\begin{figure}
\epsfxsize7cm
\hspace*{.4cm}\epsffile{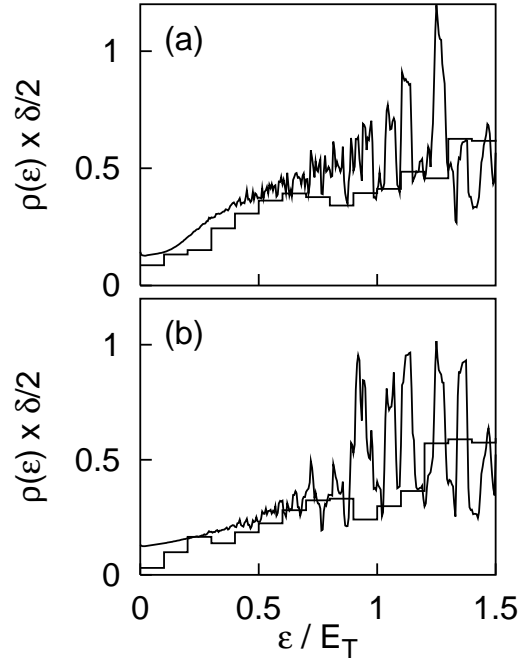}
\caption{
\narrowtext
Density of states of the billiard of Fig.\ \protect\ref{fig1}, but with
two different locations of the interface [$s=0$ in (a) and
$s=1$ in (b)]. The semiclassical prediction from Eq.\ (\protect\ref{eq:scquant}) (solid
curves) is compared with an exact quantum mechanical computation (histograms).
}
\label{fig4}
\end{figure}

The regular island is disconnected from the superconductor
when the point contact is moved to the other end of the billiard
(at $s=0$, where the separation of the circles is smallest). 
The gap in the chaotic partial density of states is
reduced to a lesser degree than before, see Fig.\ \ref{fig4}(a).
Excitations localised in the regular island
give a constant background
contribution
$2/\delta_2=2mV_2/(2\pi\hbar)^2$.
If the point contact is placed between these two extreme positions (at
$s=1$), the regular  regions of phase space dominate the low-energy
behaviour of $\rho(\varepsilon)$. 
Instead of an excitation gap we observe a smoothly and slowly
increasing density of states, see Fig.\ \ref{fig4}(b). 
The histograms in Fig.\ \ref{fig4} fall systematically below
the semiclassical prediction. We attribute this discrepancy
to the constant background contribution in the semiclassical result,
which should vanish at small $\varepsilon$ because of quantum
mechanical tunnelling through the dynamical barrier between regions 1 and 2. 
This source of error is absent in Fig.\ \ref{fig3}, because there
all regions are directly coupled to the superconductor.

In summary, we have found that the superconductor proximity effect
in ballistic systems depends sensitively on the morphology of the
classical phase space. The excitation spectrum at low energies
can be described in an intuitively appealing way by means
of effective Thouless energies and level spacings for the regular and
chaotic regions of phase space.
If the coupling to the regular regions is maximal, the excitation
spectrum exhibits an excitation gap that is much smaller than the gap of
a fully chaotic system. Measurement of such a reduced gap would provide
a unique insight into the effect of a mixed classical phase space on
superconductivity.

This work was supported by the European Community (Program for the Training
and Mobility of Researchers) and by the Dutch Science Foundation
NWO/FOM.

\ecols

\begin{thebibliography}{10}
\bibitem{numerics}
Discrepancies between exact numerical and semiclassical results
for the rectangular billiard found in Ref. \cite{Melsen} are actually within
statistical error bars.
We re-did the calculation for a greatly extended ensemble
of billiards (variations in Fermi wavelength at fixed shape)
and found good agreement.
\bibitem{tails}
One finds algebraic tails in $P_i(L)$ also for chaotic
parts of phase space,
due to trapping of orbits in the vicinity of borders to regular regions
\cite{Bauer}.
Since these tails appear only for lengths
$L\gg L_{T,i}$, they are irrelevant for the present study.
\bibitem{subdivision}
The chaotic region can be subdivided further into three parts
separated by
weakly impenetrable dynamical barriers (penetration time
larger than $L_T/v_F$).
The qualitative description is improved if one assigns to each part
its own set of effective quantities $E_{T,i}$, $\delta_i$.
\end{thebibliography}
\end{document}